\documentclass[aps,prl,reprint,groupedaddress,amsmath,amssymb,]{revtex4-1}

\usepackage{graphicx}
\usepackage{dcolumn}
\usepackage{bm}
\usepackage{multirow}
\usepackage{threeparttable}

\begin{document}

\preprint{APS/123-QED}

\title{Measurement of the $^{18}$Ne($\alpha$,p$_0$)$^{21}$Na reaction cross section in the burning energy region for X-ray bursts}

\author{P.J.C. Salter,$^1$ M. Aliotta,$^{1,*}$ T. Davinson,$^1$ H. Al Falou,$^2$ A. Chen,$^2$ B. Davids,$^2$ B.R. Fulton,$^3$ N. Galinski,$^{2,4}$ D. Howell,$^{2,4}$ G. Lotay,$^1$ P. Machule,$^2$ A. StJ. Murphy,$^1$ C. Ruiz,$^2$ S. Sjue,$^2$ M. Taggart,$^3$ P. Walden,$^2$ P.J. Woods$^1$}

\affiliation{$^1$SUPA, School of Physics and Astronomy, University of Edinburgh, Edinburgh EH9 3JZ, United Kingdom}
\affiliation{$^2$TRIUMF, Vancouver, British Columbia V6T 2A3, Canada}
\affiliation{$^3$Department of Physics, University of York, York YO10 5DD, United Kingdom}
\affiliation{$^4$Department of Physics, Simon Fraser University, Burnaby, British Columbia, Canada}

\date{\today}

\begin{abstract}
The $^{18}$Ne($\alpha$,p)$^{21}$Na reaction provides one of the main HCNO-breakout routes into the {\em rp}-process in X-ray bursts. The $^{18}$Ne($\alpha$,p$_0$)$^{21}$Na reaction cross section has been determined for the first time in the Gamow energy region for peak temperatures T$\sim$2GK by measuring its time-reversal reaction $^{21}$Na(p,$\alpha$)$^{18}$Ne in inverse kinematics. The astrophysical rate for ground-state to ground-state transitions was found to be a factor of 2 lower than Hauser-Feshbach theoretical predictions. 
Our reduced rate will affect the physical conditions under which breakout from the HCNO cycles occurs via the $^{18}$Ne($\alpha$,p)$^{21}$Na reaction.

\begin{description}
\item[PACS numbers]{25.60.-t, 26.30.Ca, 25.45.Hi}
\end{description}
\end{abstract}


\maketitle

Type I X-ray bursters (XRBs) exhibit brief recurrent bursts of intense X-ray emission and represent a frequent phenomenon in our Galaxy. Recent observations from space-borne X-ray satellites (BeppoSAX, RXTE, Chandra and XMM Newton) have provided a great wealth of data and have marked a new era in X-ray astronomy. Yet, to fully exploit these observations, similar advances in our understanding of the nuclear reactions responsible for the bursts are required. XRBs are driven by a thermonuclear runaway on the surface of a neutron star that accretes H- and He-rich material from a less evolved companion in a semi-detached binary system \cite{sch06}. Depending on the mass accretion rate, high enough temperatures and densities can be achieved that trigger hydrogen burning through the hot, $\beta$-limited CNO cycles (HCNO) and the subsequent ignition of the triple-$\alpha$ process. However, the thermonuclear runaway requires a breakout from the HCNO cycle and the ignition of the rapid-proton capture process ({\em{rp}}-process) at peak temperatures T$\simeq$ 1-2 GK. 
The $^{18}$Ne($\alpha$,p)$^{21}$Na reaction is believed to provide the main breakout route at T$\geq$0.8GK and $\rho\geq$10$^{5}$ g/cm$^{3}$ \cite{wie99}, but the actual physical conditions at which the breakout occurs depend critically on the accurate knowledge of the $^{18}$Ne($\alpha$,p)$^{21}$Na reaction rate. A direct investigation of this important reaction is severely hampered by the low intensity ($\leq 10^6$pps) of radioactive $^{18}$Ne beams presently available and by the further complications associated with the use of a $^4$He gas target. Thus, the only two direct measurements available to date extend to minimum energies of E$_{\rm cm}$= 2.0 MeV \cite{bra99} and E$_{\rm cm}$= 1.7 MeV \cite{gro02}. These are still too high compared to the energy region E$_{\rm cm}\leq$1.5 MeV of interest for HCNO breakout in X-ray bursts.\par

The first theoretical estimates of the $^{18}$Ne($\alpha$,p)$^{21}$Na reaction rate \cite{goe95} were based on sparse experimental information on the level structure of the compound nucleus $^{22}$Mg above the $\alpha$-particle threshold at 8.14 MeV. Inspection of the structure of the mirror nucleus $^{22}$Ne reveals a high level density at comparable excitation energies and suggests that a statistical approach might provide a reliable estimate of the reaction rate. However, only natural-parity states in $^{22}$Mg can be populated by the $^{18}$Ne+$\alpha$ channel and the resulting level density may be significantly smaller than required for a statistical approach. Thus, a range of experimental studies have focused on the investigation of $\alpha$-unbound natural-parity states in $^{22}$Mg using reactions such as $^{12}$C($^{16}$O,$^6$He)$^{22}$Mg, $^{25}$Mg($^3$He,$^6$He)$^{22}$Mg, $^{24}$Mg($^4$He,$^6$He)$^{22}$Mg \cite{che01, cag02, ber03}. More recently, several $\alpha$-unbound states were identified using the $^{24}$Mg(p,t)$^{22}$Mg reaction and the improved precision on the measured excitation energies resulted in smaller uncertainties on the $^{18}$Ne($\alpha$,p)$^{21}$Na reaction rate \cite{mat09}. Yet, a comparison of all the reaction rates currently available shows discrepancies of up to several orders of magnitude both below and above T $\sim $1GK (see \cite{mat09} and references therein). In addition, it remains unclear whether the Hauser-Feshbach statistical calculations provide a reliable estimate of the $^{18}$Ne($\alpha$,p)$^{21}$Na rate in the whole temperature region relevant to nucleosynthesis in X-ray bursts. \par

In this Letter, we report on the results of a time-reversal investigation to determine the $^{18}$Ne($\alpha$,p$_0$)$^{21}$Na reaction cross section at E$_{\rm cm} \simeq$1.2-2.6 MeV. 
This is the first measurement of this critical reaction in the Gamow energy region for a maximum peak temperature T$\simeq$ 2GK (for a given reaction, the Gamow peak energy and width vary with temperature as T$^{2/3}$ and T$^{5/6}$, respectively). As only ground-state to ground-state transitions in the $^{18}$Ne($\alpha$,p)$^{21}$Na reaction can be accessed from the $^{21}$Na(p,$\alpha$)$^{18}$Ne time-reversal reaction, the inferred cross section represents a lower limit to the total $^{18}$Ne($\alpha$,p)$^{21}$Na reaction cross section and further investigations will be needed to assess any additional inelastic contributions (see \cite{he10} for such a study in the case of $^{14}$O($\alpha$,p)$^{17}$F). It should be noted, furthermore, that given the non-zero spins of the interacting nuclei in the entrance channel (J$^\pi = 3/2^+$ and J$^\pi = 1/2^+$ for $^{21}$Na and proton, respectively) both natural- and non-natural-parity states can be populated in $^{22}$Mg. However, detection of $^{18}$Ne nuclei in their ground state ensures that only natural-parity states in $^{22}$Mg, {\em i.e.} those of astrophysical interest, have been populated by the inverse 
$^{21}$Na(p,$\alpha$)$^{18}$Ne reaction. \par

The time-reversal measurement was carried out at the ISAC II facility (TRIUMF) in inverse kinematics using a radioactive $^{21}$Na beam incident on (CH$_{2}$)$_{n}$ targets (thicknesses 311 and 550 $\mu$g/cm$^2$). Six beam energies were explored in the region 4.12 MeV/A $\leq$ E$_{\rm beam} \leq$5.48 MeV/A. Typical beam intensities were limited to a maximum of 10$^{6}$ pps in order to keep the Rutherford scattering yield in the most forward detectors (see below) at an acceptable rate. Heavy ions and $\alpha$ particles were detected in coincidence using an array of Double Sided Silicon Strip Detectors (DSSSDs). Detection of $^4$He ions was primarily achieved with a $\Delta$E-E telescope consisting of two MSL \cite{micron} type S2 detectors (65 $\mu$m and 500 $\mu$m thick, respectively, each segmented into 48 annular front strips and 16 rear sectors) covering a laboratory angular range of $\theta_{\alpha}= 7^o-19^o$ (the maximum emission angle for $\alpha$ particles was less than $20^o$ at all energies investigated). Detection of $^{18}$Ne ions was  achieved with a $\Delta$E-E telescope using a ``CD'' detector consisting of 4 quadrants (each segmented into 16 front strips and 24 rear sectors) and a ``PAD'' detector, consisting of 4 unsegmented quadrants \cite{micron}. The CD-PAD (35 $\mu$m and 1500 $\mu$m thick, respectively) covered a laboratory angular range $\theta_{\rm HI}= 1.6^o-6.6^o$ (the maximum emission angle for $^{18}$Ne ions was less than $\sim 5^o$ at all energies investigated). The coincidence detection efficiency was between 13$\%$ and $27\%$, depending on beam energy, as determined by a Monte Carlo simulation assuming an isotropic distribution in the center of mass \cite{sal11}. \par

\begin{figure}
  \includegraphics[width=\linewidth]{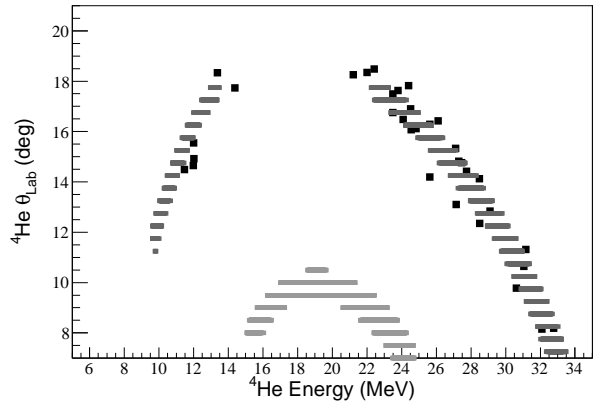}
  \caption{\label{fig:alpha-kin}Simulated alpha-particle kinematics curves ($\theta$ {\em vs.} E) superimposed on experimental data (black squares) from the $^{21}$Na(p,$\alpha$)$^{18}$Ne reaction at E$_{\rm beam}$= 5.476 MeV/A. The upper (lower) locus corresponds to reactions forming $^{18}$Ne in its ground- (first-excited) state.}
\end{figure}

The energy calibration of the detectors was performed independently for each strip using a mixed three-peak $\alpha$ source. In addition, the Rutherford elastic scattering of $^{21}$Ne off $^{12}$C at E$_{\rm lab}$=5.357 MeV/A was  used as a fourth calibration point for the CD quadrants only. 
Good events, corresponding to $^4$He and $^{18}$Ne ions detected in coincidence, were extracted from the raw data by imposing appropriate conditions of two-body co-planarity, identification of $^4$He and $^{18}$Ne kinematics loci, and total energy reconstruction. An example of the measured and simulated $\alpha$-particle kinematics loci for the $^{21}$Na(p,$\alpha$)$^{18}$Ne reaction at E$_{\rm beam}$= 5.476 MeV/A is shown in Figure \ref{fig:alpha-kin}. The upper and lower loci correspond to $^{18}$Ne nuclei being left in their ground state or first excite state (E$_x$=1.89 MeV), respectively. This latter channel is open at beam energies 5.476 and 4.910 MeV/A (and closed at the other energies investigated).  No events were observed in the lower locus at either beam energy, thus indicating that the inelastic $^{21}$Na(p,$\alpha$)$^{18}$Ne$^*$ channel is not strongly populated in the reaction. Experimental data were found to be in excellent agreement with simulations at all beam energies.\par

The $^{21}$Na(p,$\alpha$)$^{18}$Ne cross section was determined from the measured coincident yields under the assumption of a thin-target condition at each investigated energy, using the following equation:
$$\sigma(E_{\rm eff}) = \frac{Y}{N_bN_t\zeta\tau}$$
where $E_{\rm eff}$ is the effective beam energy at mid-target, $Y$ is the measured $^{21}$Na(p,$\alpha$)$^{18}$Ne coincident yield, $N_b$ the total number of incident particles, $N_t$ the number of $^1$H nuclei per unit area in the (CH$_2$)$_n$ target, $\zeta$ the coincident detection efficiency, and $\tau$ is the data acquisition (DAQ) live time (typically 65\%). The number of incident particles was calculated from the yield of Rutherford scattered $^{21}$Na ions off $^{12}$C in the target as detected in the innermost strips of the CD detectors, {\em i.e} well within the grazing angle at each beam energy.  

A summary of experimental yields and cross sections for each of the six beam energies investigated is given in Table \ref{tab:(p,a)-yields}. 
\begin{table*}[htb]
  \caption{\label{tab:(p,a)-yields}Summary of experimental results for each of the six beam energies investigated.}
  \begin{ruledtabular}
  \begin{threeparttable}
    \begin{tabular}{ c c c c c c c c }
E$_{\rm beam}$ & E$^{\rm eff}_{\rm{cm}}$ (p,$\alpha$) & E$^{\rm eff}_{\textrm{cm}}$($\alpha$,p) & $^{4}$He+$^{18}$Ne & $\zeta$\tnote{c} & N$_{\rm b}$\tnote{d} & $\sigma$(p,$\alpha$) & $\sigma$($\alpha$,p)\\
(MeV/A) & (MeV) & (MeV) & Yield & ($\%$) & particles & (mb) & (mb)\\
\hline
5.476\tnote{a} & 5.21$\pm$0.06 & 2.57$\pm$0.06 & 33$\pm$6 & 19 & $(2.9\pm0.3)\times10^{10}$ & 0.35$\pm$0.06 & 1.7$\pm$0.3 \\
4.910\tnote{b} & 4.61$\pm$0.12 & 1.97$\pm$0.12 & 8$^{+3.3}_{-2.7}$ & 27 & $(3.2\pm0.3)\times10^{10}$ & $\left(3.0^{+1.2}_{-1.0}\right)\times10^{-2}$ & 0.17$^{+0.07}_{-0.06}$ \\
4.642\tnote{a} & 4.40$\pm$0.07 & 1.76$\pm$0.07 & 23$\pm$5 & 27 & $(7.1\pm0.7)\times10^{11}$ & $\left(5.3\pm1.1\right)\times10^{-3}$ & $\left(3.1\pm0.6\right)\times10^{-2}$ \\
4.619\tnote{b} & 4.32$\pm$0.12 & 1.68$\pm$0.12 & 16$^{+5}_{-4}$ & 25 & $(5.0\pm0.5)\times10^{11}$ & $\left(3.8^{+1.1}_{-0.9}\right)\times10^{-3}$ & $\left(2.3^{+0.7}_{-0.5}\right)\times10^{-2}$ \\
4.310\tnote{b} & 4.02$\pm$0.13 & 1.38$\pm$0.13 & 4$^{+2.8}_{-1.7}$ & 16 & $(1.47\pm0.15)\times10^{12}$ & $\left(5.6^{+3.9}_{-2.3}\right)\times10^{-4}$ & $\left(3.8^{+2.7}_{-1.6}\right)\times10^{-3}$ \\
4.120\tnote{b} & 3.83$\pm$0.13 & 1.19$\pm$0.13 & 2$^{+2.3}_{-1.3}$ & 13 & $(6.9\pm0.7)\times10^{12}$ & $\left(7.4^{+8.3}_{-4.6}\right)\times10^{-5}$ & $\left(5.5^{+6.2}_{-3.5}\right)\times10^{-4}$ \\
    \end{tabular}
\begin{tablenotes}
\item[{a}] 311$\mu$g/cm$^2$ (CH$_2$)$_n$ target
\item[{b}] 550$\mu$g/cm$^2$ (CH$_2$)$_n$ target
\item[{c}] as determined by Monte Carlo simulation
\item[{d}] as determined by Rutherford elastic scattering of $^{21}$Na beam off $^{12}$C nuclei in the target (see text).
\end{tablenotes}
\end{threeparttable}
  \end{ruledtabular}
\end{table*}
Note that the errors in the effective interaction energy represent half the target thickness (in the center-of-mass system) at  each beam energy. The errors in the yields and cross sections are statistical only and were calculated using Poisson statistics for the two highest yields and using the Feldman-Cousins method \cite{fel98} for low statistics for all other yields (68$\%$ confidence level and zero background assumption). 
The estimated systematic uncertainty of the cross section is  $\sim$16\% and is dominated by uncertainties in the number of projectile nuclei (8$\%$ at all beam energies) and in the number of target nuclei (about 8$\%$, mostly due to the uncertainty in the  energy loss calculations by SRIM2008 \cite{SRIM}). Changes to the detection efficiency due to non-isotropic distributions (up to $l$=3) were also explored by Monte Carlo simulations. Deviations from the isotropic case amount to at most 20\% at the highest beam energies and to at most 50-60\% at the two lowest beam energies (depending on the actual $l$ value) and are therefore comparable to, or smaller than, the quoted statistical uncertainties.

\begin{figure}
  \includegraphics[width=\linewidth]{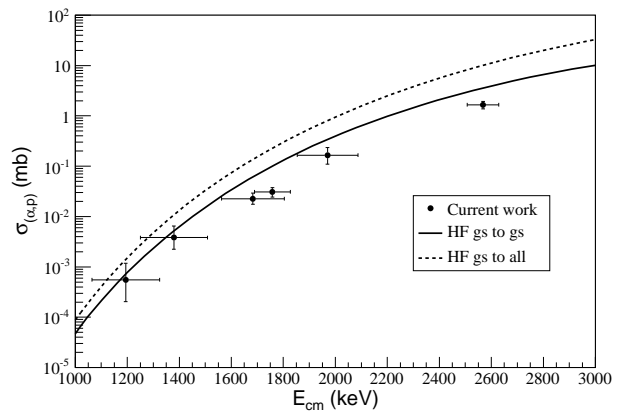}
  \caption{\label{fig:cross-section} Experimental $^{18}$Ne($\alpha$,p$_0$)$^{21}$Na reaction cross section (black dots) as a function of E$_{\textrm{cm}}^{(\alpha,p)}$. Predictions based on the Hauser-Feshbach calculations \cite{rau01} for ground-state to ground-state transitions (full line) and ground-state to all-states transitions (dashed line) are also shown for comparison.}
\end{figure}

The $^{18}$Ne($\alpha$,p$_0$)$^{21}$Na reaction cross section was inferred by using the principle of detailed balance \cite{ili07} and is shown in Figure \ref{fig:cross-section} as a function of ($\alpha$,p) center-of-mass energy. The figure also shows the theoretical predictions based on Hauser-Feshbach calculations \cite{rau01} for ground-state to ground-state transitions (hereafter HF$_{\rm gs}$) and ground-state-to-all-states transitions (HF$_{\rm all}$) for the $^{18}$Ne($\alpha$,p)$^{21}$Na reaction. Since our data only provide a lower limit to the cross section, the comparison with theoretical predictions is made only with HF$_{\rm gs}$. Surprisingly, a good agreement is found for the two lowest energy points, while a discrepancy of up to a factor of 2 is observed at the highest measured energy. This is contrary to expectations, as lower energies correspond to lower excitation energies, and therefore lower level densities, in the compound nucleus. The reason for such a trend is at present not understood. 

\begin{figure}
  \includegraphics[width=\linewidth]{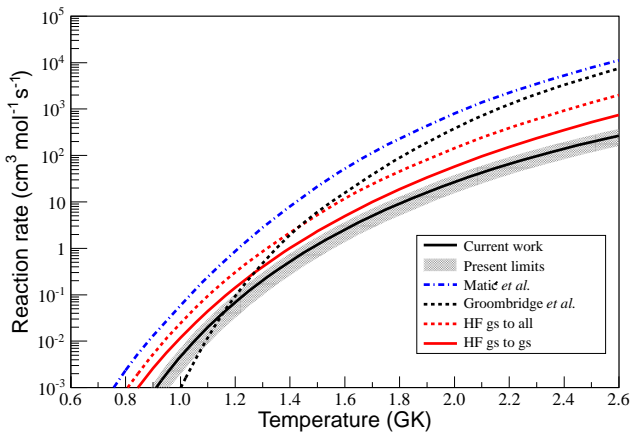}
  \caption{\label{fig:reaction-rate}(Colour online) Astrophysical $^{18}$Ne($\alpha$,p$_0$)$^{21}$Na reaction rate from the present work and associated uncertainties (solid black line and shaded area). Theoretical predictions based on Hauser-Feshbach formalism are also shown for comparison (red lines). The total stellar rates from \cite{gro02} (black dotted line) and \cite{mat09} (blue dash-dotted line) are larger than the present rate by up to a factor of 25 and 40, respectively, at T=2.4GK. This temperature is  chosen as an example to illustrate quantitative differences between the various rates. }
\end{figure}

The astrophysical $^{18}$Ne($\alpha$,p$_0$)$^{21}$Na reaction rate as a function of temperature was calculated by numerical integration of our experimental cross sections using the {\em exp2rate} code by T. Rauscher \cite{rauscher}. Rate values, obtained as the arithmetic mean between the low and high limits associated with the uncertainties on the cross section data, are given in Table \ref{tab:reaction-rate}. As shown in Figure \ref{fig:reaction-rate}, our reaction rate agrees well with the overall energy dependence of the HF$_{\rm gs}$ rate but is typically a factor of $\sim$2-3 lower in the whole temperature region (T=1.0-2.4 GK) that corresponds to our measured energy range.  However, a comparison with the rate from the direct measurement of Groombridge {\em et al.} \cite{gro02} reveals significant discrepancies both in energy dependence and magnitude, with a disagreement of up to a factor of $\simeq$ 25 at T=2.4GK. Our rate is also between a factor of about 10-40 lower than the rate deduced using an indirect approach by  Mati\'{c} {\em et al.} \cite{mat09} for 0.9 GK$\leq$T$\leq$2.4 GK. 
It should be pointed out that such discrepancies would persist (albeit to a lesser extent) even by allowing for inelastic contributions to our rate because, according to HF predictions, these latter would increase our rate by at most a factor of $\sim$3 (see Fig. \ref{fig:cross-section} and \ref{fig:reaction-rate}). The origin of these large discrepancies is not clear but we note that the resonance widths used in \cite{gro02,gro02b} imply unphysically large alpha-decay partial widths. Similarly, the rate in \cite{mat09} is subject to large uncertainties in the choice of resonance strength values, which are not determined experimentally. Outside of the temperature range shown in Figure \ref{fig:reaction-rate} our rate can be extrapolated but its reliability decreases drastically, especially for T $\leq$ 0.95 GK and thus a direct comparison with the rate of \cite{che01} is not possible. 

\begin{table}
  \caption{\label{tab:reaction-rate}Astrophysical $^{18}$Ne($\alpha$,p$_0$)$^{21}$Na reaction rate as a function of temperature, calculated by numerical integration of our cross-section data. The rate is taken as the arithmetic mean of low and high limits associated with the uncertainties on the cross sections.}
  \begin{ruledtabular}
    \begin{tabular}{ c c c c }
Temperature & \multicolumn{3}{c}{N$_{A} <\sigma\upsilon>$ (cm$^{3}$ mol$^{-1}$ s$^{-1}$)} \\
T$_{9}$ (K) & low limit & high limit & arithmetic mean \\
\hline
0.95 & 8.5$\times10^{-4}$ & 3.2$\times10^{-3}$ & (2.0 $\pm$ 1.2)$\times10^{-3}$\\
1.05 & 4.4$\times10^{-3}$ & 1.6$\times10^{-2}$ & (9.9 $\pm$ 5.6)$\times10^{-3}$\\
1.15 & 1.7$\times10^{-2}$ & 5.9$\times10^{-2}$ & (3.8 $\pm$ 2.1)$\times10^{-2}$\\
1.25 & 5.8$\times10^{-2}$ & 1.8$\times10^{-1}$ & (1.2 $\pm$ 0.6)$\times10^{-1}$\\
1.35 & 1.6$\times10^{-1}$ & 5.0$\times10^{-1}$ & (3.3 $\pm$ 1.7)$\times10^{-1}$\\
1.45 & 4.1$\times10^{-1}$ & 1.2 & (8.0 $\pm$ 3.9)$\times10^{-1}$\\
1.55 & 9.4$\times10^{-1}$ & 2.6 & 1.8 $\pm$ 0.8\\
1.65 & 2.0 & 5.3 & 3.6 $\pm$ 1.7\\
1.75 & 3.8 & 9.9 & 6.9 $\pm$ 3.1\\
1.85 & 6.9 & 1.7$\times10^{+1}$ & (1.2 $\pm$ 0.5)$\times10^{+1}$\\
1.95 & 1.2$\times10^{+1}$ & 3.0$\times10^{+1}$ & (2.1 $\pm$ 0.9)$\times10^{+1}$\\
2.05 & 2.0$\times10^{+1}$ & 4.8$\times10^{+1}$ & (3.4 $\pm$ 1.4)$\times10^{+1}$\\
2.15 & 3.1$\times10^{+1}$ & 7.5$\times10^{+1}$ & (5.3 $\pm$ 2.2)$\times10^{+1}$\\
2.25 & 4.7$\times10^{+1}$ & 1.1$\times10^{+2}$ & (8.0 $\pm$ 3.3)$\times10^{+1}$\\
2.35 & 6.9$\times10^{+1}$ & 1.6$\times10^{+2}$ & (1.2 $\pm$ 0.5)$\times10^{+2}$\\
2.45 & 9.9$\times10^{+1}$ & 2.3$\times10^{+2}$ & (1.7 $\pm$ 0.7)$\times10^{+2}$\\
    \end{tabular}
  \end{ruledtabular}
\end{table}

Based on our results, we can infer that the breakout from the HCNO cycle via the $^{18}$Ne($\alpha$,p)$^{21}$Na reaction is delayed and occurs at higher temperatures than previously predicted. However, more detailed conclusions require full hydro-dynamical model calculations for X-ray bursts that are beyond the scope of this Letter.\par

In summary, the $^{18}$Ne($\alpha$,p)$^{21}$Na reaction plays a crucial role in type I X-ray bursts as it provides a breakout route from the HCNO cycle into the {\em rp-}process that triggers the thermonuclear explosion responsible for the burst phenomenon. Using the detailed balance theorem, we have determined the $^{18}$Ne($\alpha$,p$_0$)$^{21}$Na reaction cross section by measuring its time-reversed reaction at the ISAC-II facility, TRIUMF. The measurement covered the energy region E$_{\rm cm}(\alpha$,p)=1.19-2.57 MeV with E$_{\rm cm}(\alpha$,p)=1.19 MeV being the lowest energy measured to date and, for the first time, within the Gamow energy region of this reaction in X-ray bursts.
Our results indicate that a breakout from the HCNO via the $^{18}$Ne($\alpha$,p)$^{21}$Na reaction should occur at higher temperatures than previously assumed. 

\begin{acknowledgments}
The authors wish to thank the ISAC operation and technical staff at TRIUMF, as well as C. Marchetta (LNS, Catania) and P. Demaret (LLN, Belgium) for the preparation of targets. The UK authors acknowledge support by STFC.
\end{acknowledgments}

$^*$ Corresponding author: m.aliotta@ed.ac.uk
\bibliography{s1103paper}

\end{document}